\documentclass{eusflat2021}

\usepackage{xcolor}
\usepackage{amsmath}
\usepackage{amssymb} 
\usepackage{amsfonts}
\usepackage{eucal}
\usepackage{bm}

\title{\bf Incremental Learning and State-Space Evolving Fuzzy Control \\ of Nonlinear Time-Varying Systems with Unknown Model}

\author{$^*$\textbf{Daniel Leite}$^a$ \and \textbf{Pedro Coutinho}$^b$ \and Iury Bessa$^{b,c}$  \and Murilo Camargos$^d$    \\ \and \textbf{Luiz A. Q. Cordovil Junior.}$^b$ \and $^*$\textbf{Reinaldo Palhares}$^e$ \\
$^a$Department of Automatics, Federal University of Lavras, Brazil. \email{daniel.leite@ufla.br} \\
$^b$Graduate Program in Electrical Engineering, Federal University of Minas Gerais, Brazil.\\\email{{coutinho.p92,luiz.cordovil05}@gmail.com} \\
$^c$Department of Electricity, Federal University of Amazonas, Brazil. \email{iurybessa@ufam.edu.br}\\
$^d$School of Applied Mathematics, Fundação Getulio Vargas, Rio de Janeiro, Brazil. \email{murilo.camargosf@gmail.com} \\
$^e$Department of Electronics Engineering, Federal University of Minas Gerais, Brazil. \email{rpalhares@ufmg.br} }

\begin{document}

\maketitle

\begin{abstract}

\vspace{-5pt}

We present a method for incremental modeling and time-varying control of unknown nonlinear systems. The method combines elements of evolving intelligence, granular machine learning, and multi-variable control. We propose a \textbf{S}tate-\textbf{S}pace \textbf{F}uzzy-set-\textbf{B}ased \textbf{e}volving \textbf{M}odeling (SS-FBeM) approach. The resulting fuzzy model is structurally and parametrically developed from a data stream with focus on memory and data coverage. The fuzzy controller also evolves, based on the data instances and fuzzy model parameters. Its local gains are redesigned in real-time -- whenever the corresponding local fuzzy models change -- from the solution of a linear matrix inequality problem derived from a fuzzy Lyapunov function and bounded input conditions. We have shown one-step prediction and asymptotic stabilization of the Henon chaos. 

\end{abstract}

\vspace{-9pt}

\section{Introduction}

\vspace{-2pt}

\subsection{Contextualization}

\vspace{-1pt}

Evolving fuzzy systems (eFS) \cite{Leite2020} are universal approximators whose parameters and rule-based structure are updated from never-ending data streams, potentially subject to changes. eFS have been effectively employed in systems identification \cite{Saso}, filtering~\cite{Pires2020}, prediction \cite{Cordovil2019} \cite{Leite1}, missing data handling \cite{Garcia2020}, classification \cite{Andonovski} \cite{Skr}, image recognition \cite{Gu}, fault detection \cite{Kharrat} \cite{lemos2013adaptive}, fault prognostics \cite{Camargos2020} \cite{Camargos2}, and robust control \cite{Leite2} \cite{Precup}, to mention some. 

Of concern to this paper, fuzzy control has long been recognized as a prominent tool to handle complex nonlinear systems \cite{Nguyen2019}. If the equations that describe a dynamical system are known, an exact fuzzy representation can be derived through the Sector Nonlinearity method \cite{Tanaka1}. Thus, the accurateness of the assumptions in obtaining nonlinear equations is the sole aspect that restricts the performance of fuzzy model-based control (or any other nonlinear controller) on a physical system to be similar to the performance requested and observed in computer simulations. Data-stream-driven fuzzy control comes into play as a key approach when the system equations are unknown, and/or nonstationary -- being the latter a clear drawback of any offline model-based and model-free control method since the initial assumptions change unpredictably over time. In particular, eFS tools allow to model, analyze, and control nonlinear time-varying systems with unknown dynamics based only on a data stream \cite{Leite2}.

Evolving granular computing \cite{phdthesisL} is a general-purpose online learning framework, i.e., a family of algorithms and methods to construct classifiers, regressors, predictors, and controllers in which any aspect of a problem may assume a non-pointwise (e.g., interval, fuzzy, rough, statistical) uncertain characterization, including data, parameters, attributes, learning equations, covering regions \cite{Garcia2020} \cite{phdthesisL} \cite{Wang2019}. In particular, we have proposed a state-space variety of a granular eFS known as Fuzzy-set-Based evolving Modeling (FBeM) \cite{Leite1}, and a model-based control design method that guarantees Lyapunov stability and bounded inputs to the closed-loop evolving granular system.

Although FBeM is suitable to fuzzy and interval data processing, and outputs a granule that encloses the actual data, in this study, FBeM notices a dynamic system from numerical data only. We are interested in numerical estimates, and fuzzy model parameters to support control design. As an SS-FBeM model may change at any time step, the corresponding fuzzy controller is locally redesigned from the solution a relaxed linear matrix inequality (LMI). We couple SS-FBeM with a fuzzy Lyapunov function and bounded input condition to obtain fuzzy control gains that guarantee closed-loop stability and avoid actuator saturation. We demonstrate the efficacy of the SS-FBeM predictor and associated controller using a deterministic chaotic nonlinear system, known as Henon map \cite{Henon}. 

Broadly speaking, we have achieved stabilization of unknown, nonlinear and nonstationary, systems using no \textit{a priori} information, and fully-autonomous incremental granular learning. We visualize applications on secure communication, such as in smart IoT and cyber-physical systems; suppression of interference and artifacts in cardiac, electroencephalogram, and speech signals; and control of switched and time-varying systems in general, to mention some.

\color{black}

\vspace{-1pt}

\subsection{Formal problem statement} \label{sec:problem}

\vspace{-1pt}

Consider a general nonlinear and time-varying system in the discrete form,

\vspace{-16pt}

\begin{eqnarray}
\textbf{x}(k+1) &=& f\left(\textbf{x}(k),\textbf{u}(k),k\right) \nonumber \\
\textbf{y}(k+1) &=& h\left(\textbf{x}(k),k\right), \label{model1}
\end{eqnarray}

\vspace{-4pt}

\noindent in which $\textbf{x}(k) \in R^n$, $\textbf{u}(k) \in R^m$, and $\textbf{y}(k) \in R^p$ are the state, input and output vectors at time step $k$. The model (\ref{model1}) is defined by the maps $f$ and $h$. We consider all states accessible. Then, $\textbf{y}(k) = \textbf{x}(k)$, and the output equation is omitted. The goal is to design a state feedback control law of the form

\vspace{-13pt}

\begin{eqnarray}
\textbf{u}(k) = g\left(\textbf{x}(k),k\right) \label{law}
\end{eqnarray}

\vspace{-5pt}

\noindent such that a fixed point of the unknown map $f$, namely the origin, is stabilized.

Since $f$ has an unknown model, the control law, $g$, can only be designed for $f'$ -- a model of $f$. An evolving granular fuzzy model $f'$ is considered. As $f$ is time varying in general, the model $f'$ must be provided with mechanisms for learning from a sequence of data $\textbf{x}(k)$, $k = 1, ...$. In other words, $f'$ must be evolved over time to track the behavior of $f$. It is expected that $f$ is stabilized by the control $\textbf{u}(k)$ if $f'$ is a good approximation of $f$, and $g$ is an appropriate control law.

A solution $\textbf{x}^*(k)$ of \eqref{model1}, with initial condition $\textbf{x}^*(0)$, is called chaotic if it is Lyapunov unstable and all the solutions starting from some neighborhood of $\textbf{x}^*(0)$ are bounded on $(-\infty,\infty)$. Stabilizing $\textbf{x}^*(k)$ means to drive the states $\textbf{x}(k)$ to $\textbf{x}^*(k)$, i.e.,

\vspace{-13pt}

\begin{eqnarray}
\lim_{k\rightarrow\infty} (\textbf{x}(k) - \textbf{x}^*(k)) = 0.
\end{eqnarray}

\vspace{-5pt}

\begin{figure}[ht]
\begin{center}
\includegraphics[width=.99\columnwidth]{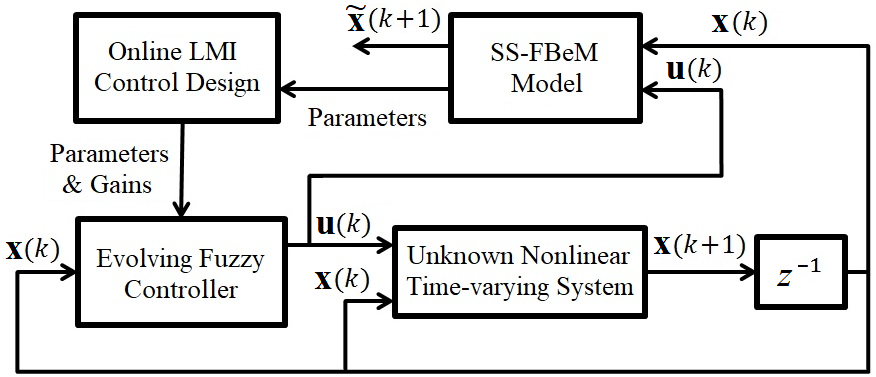}
  \caption{\label{figure1} State-space evolving fuzzy control system.}
\end{center}
\end{figure}

Figure \ref{figure1} shows the closed-loop system with the evolving state-space fuzzy model, $f'$, and controller, $g$, we are interested. In the figure, $\widetilde{\textbf{x}}(k+1)$ is an estimate of the state at $k+1$; and $z^{-1}$ is the delay operator. The \textit{nonlinear system with unknown model}, $f$, is a general definition that extends to objects, machines, and a variety of virtual and real-world systems that can be controlled. We notice $f$ from a data stream.

\vspace{-8pt}

\section{State-Space Fuzzy-set-Based evolving Model: SS-FBeM} \label{sec:ss_fbem}

A state-space variety of the fuzzy-set-based evolving modeling, FBeM, approach by \cite{Leite1} is introduced. The SS-FBeM learning algorithm underlines data coverage and memory of the past. The state-space fuzzy model aims to assist online LMI control design.

\vspace{-1pt}

\subsection{Evolving fuzzy modeling}

\vspace{-1pt}

We give an incremental learning algorithm to build a fuzzy rule-based model with state-space consequent from a data stream produced by a dynamical system. The dynamical system's difference equations are unknown, nonlinear, and time-varying in general. A finite number of past states $\textbf{x}(k), \textbf{x}(k-1), ..., \textbf{x}(k-d)$; control inputs $\textbf{u}(k), \textbf{u}(k-1), ..., \textbf{u}(k-v)$; and other external variables can be considered as antecedent terms of rules. Nonetheless, a common assumption in model-based fuzzy control is rules of the form

\vspace{3pt}

\noindent ~~ $R^i$: IF $x_1(k)$ is $\mathcal{M}_1^i$ AND ... AND $x_{\Psi}(k)$ is $\mathcal{M}_{\Psi}^i$ 

\noindent ~~~~~~~ THEN $\textbf{x}^i(k+1) = A^i \textbf{x}(k) + B^i \textbf{u}(k)$

\vspace{3pt}

\noindent in which $\textbf{x}(k) = [x_1(k)$ ... $x_\psi(k)$ ... $x_\Psi(k)]^T$ and $\textbf{u}(k)$ = $[u_1(k)$ ... $u_\phi(k)$ ... $u_\Phi(k)]^T$; $R^i$, $i = 1, ..., c$, is a variable amount of rules. In evolving modeling, $A^i$ is a matrix of appropriate dimension with variable coefficients; $\mathcal{M}_\psi^i$, $\psi = 1, ..., \Psi$, are trapezoidal membership functions -- defined by four strictly increasing parameters, namely $\mathcal{M}_\psi^i = (l_\psi^i,\lambda_\psi^i,\Lambda_\psi^i,L_\psi^i)$ -- created and updated in light of the data being available. We assume $B^i = [1 ~ 0 ~ ... ~ 0]^T$ common to all rules, namely, the control input is applied to $x_1$, without loss of generality. Superscript $i$ on the left-hand side of the consequent equation means a local one-step estimate.

We extended the consequent matrices, and the state and control vectors, to include affine terms. Thus,

\vspace{-10pt}

\begin{eqnarray} 
~~ \widetilde{A}^i = \hspace{-1pt} \left[ \begin{array}{cc}
1 \hspace{-2pt} & \textbf{0} \\
\textbf{a}_{0}^i \hspace{-2pt} & A^i      
\end{array} \right] \hspace{-2pt} , \hspace{-2pt} ~
\widetilde{B}^i = \hspace{-1pt} \left[ \begin{array}{c}
0 \\
B^i     
\end{array} \right] \hspace{-2pt} , \hspace{-2pt} ~
\widetilde{\textbf{x}} = \hspace{-1pt} \left[ \begin{array}{c}
1 \\
\textbf{x}    
\end{array} \right] \hspace{-2pt} , \hspace{-2pt} ~
\widetilde{\textbf{u}} = \hspace{-1pt} \left[ \begin{array}{c}
u_0 \\
\textbf{u}     
\end{array} \right] \hspace{-2pt} , \nonumber
\end{eqnarray}

\vspace{-7pt}

\noindent being $\textbf{a}_{0}^i = [a_{10}^i$ ... $a_{\psi0}^i$ ... $a_{\Psi0}^i]^T$; and $u_0$ the controller offset term, which is obtained straightforwardly from the design method. Rules are rewritten as

\vspace{3pt}

\noindent ~~ $R^i$: IF $x_1(k)$ is $\mathcal{M}_1^i$ AND ... AND $x_{\Psi}(k)$ is $\mathcal{M}_{\Psi}^i$ 

\noindent ~~~~~~~ THEN $\widetilde{\textbf{x}}\hspace{1pt}^i(k+1) = \widetilde{A}^i \hspace{1pt} \widetilde{\textbf{x}}(k) + \widetilde{B}^i \hspace{1pt} \widetilde{\textbf{u}}(k)$

\vspace{3pt}

\noindent For simplicity, we omit `$(k)$' from the time-varying membership functions $\mathcal{M}_\psi^i$, and system matrices $A^i$.

The global state estimate of the fuzzy model is

\vspace{-8pt}

\begin{equation}
~~ \widetilde{\textbf{x}}(k+1) = \sum\limits_{i=1}^c \mu^{ir} \widetilde{\textbf{x}}^i(k+1); \label{outest}
\end{equation}

\vspace{-4pt}

\noindent $\mu^{ir}$ is the re-scaled activation degree of the $i$-th rule, 

\vspace{-12pt}

\begin{eqnarray}
~~ \mu^{ir} = \frac{\mu^i}{\sum\limits_{i=1}^c \mu^i}, \textrm{ so that } \mu^{ir} \geq 0 \textrm{ and } \sum_{i=1}^{c} \mu^{ir} = 1. \label{boundc}
\end{eqnarray}

\vspace{-6pt}

Rules activation degrees $\mu^{i}$, $i = 1, ..., c$, can be determined using an aggregation operator, e.g., a T-norm \cite{Beliakov1}, i.e., $\mu^i = T(\mu_1^i, ...,  \mu_\Psi^i)$, in which $\mu_{\psi}^i$ is the membership degree of $x_\psi(k)$ in $\mathcal{M}_\psi^i$. Nevertheless, we preserve the original FBeM procedure to obtain $\mu^{i}$ from a similarity measure between fuzzy objects, and from the notion of expansion region.

Let the expansion region of $\mathcal{M}_\psi^i$ be denoted by

\vspace{-12pt}

\begin{eqnarray}
E_\psi^i ~ = ~ [L_\psi^i - \rho, ~ l_\psi^i + \rho], \label{exp}
\end{eqnarray}

\vspace{-5pt}

\noindent in which $\rho$ is the maximum width that $\mathcal{M}_\psi^i, ~ \forall \psi, i$, is allowed to expand to enclose a datum $x_\psi$. In other words, $L_\psi^i - l_\psi^i \leq \rho, ~ \forall \psi, i$, at any $k$. 

Define $\mathcal{M}^i = [\mathcal{M}^i_1 ~ ... ~ \mathcal{M}^i_\Psi]^T$ as a $\Psi$-dimensional information granule drew in the state space from the cylindrical extension of its elements. $\textbf{E}^i = [E_1^i ~ ... ~ E_\Psi^i]^T$ is the expansion region of the granule $\mathcal{M}^i$.

Given a numerical instance $\textbf{x}$ at instant $k$, the activation degree of an SS-FBeM rule, say $R^i$, is $\mu^i$ = $S(\textbf{x},\mathcal{M}^i)$ if $\textbf{x} \in \textbf{E}^i$ -- being $S(.)$ a similarity measure; otherwise $\mu^i = 0$. We use

\vspace{-14pt}

\begin{eqnarray}
S(\textbf{x},\bm{\mathcal{M}}^i) &=& 1 - \frac{1}{6\Psi} \sum_{\psi=1}^\Psi ( |x_\psi - l_\psi^{i}| + 2 |x_\psi - \lambda_\psi^i| + \nonumber \\
&& + 2 |x_\psi - \Lambda_\psi^i| + |x_\psi - L_\psi^i| ). \label{sss}
\end{eqnarray}

\vspace{-4pt}

\noindent The value of $S$ equals $1$ (indicating maximum activation) if the trapezoids $(l_\psi^{i},\lambda_\psi^{i},\Lambda_\psi^{i},L_\psi^{i})$, $\forall \psi$, are degenerated in singletons, and match $\textbf{x}$. The similarity reduces linearly as $\textbf{x}$ withdraws from $\mathcal{M}^i$ in any dimension. In particular, the summatory function in Eq. (\ref{sss}) determines a Hamming-like distance, which involves only basic arithmetic operations \cite{Leite1} \cite{Cross}.

While many methods assume the activation degree $\mu^i$ of at least one rule $R^i$ to be nonzero, this is not the case in evolving environment since no fuzzy set exists \emph{a priori}. Sets and rules are created and developed gradually to cover the state space. The number of rules, $c$, increases by a unit if $\mu^i = 0, ~ \forall i$. In this case, $\mu^{c+1} = 1$, i.e., the fuzzy sets of the new rule match the data instance. Online learning is addressed next.

\vspace{-1pt}

\subsection{Online incremental learning}

\vspace{-1pt}

SS-FBeM is built from scratch. The model acquires information from a data stream in response to new behaviors and changes. We describe a learning method that deals with time-varying nonlinear systems and avoids time-consuming batch training -- common to conventional machine learning methods.

Expansion regions, $\textbf{E}^i$, \eqref{exp}, are essential to the decision on whether or not a new instance belongs to a granule $\mathcal{M}^i$. Different values of the hyperparameter $\rho \in (0,1)$ produce different granular representations of the same dynamical system. The lower the value of $\rho$, the larger the SS-FBeM structure, and the greater the details the model can seize. However, if $\rho$ tends to $0^+$, granules are not expanded; a rule is created for each instance, which causes excessive complexity. In contrast, if $\rho$ is equal to 1, a single granule covers the data, which is insufficient to nonlinear modeling and control.


An SS-FBeM rule is created whenever one or more entries of $\textbf{x}(k)$ do not belong to the expansion regions $\textbf{E}^i$ of $\bm{\mathcal{M}}^i$, $\forall i$, $i = 1, ..., c$. The new granule $\bm{\mathcal{M}}^{c+1}$ is constructed from fuzzy sets $\mathcal{M}_\psi^{c+1}$, $\psi = 1,...,\Psi$, whose parameters match $\textbf{x}(k)$, i.e.,

\vspace{-13pt}

\begin{equation}
\mathcal{M}_\psi^{c+1} = (l_\psi, \lambda_\psi, \Lambda_\psi, L_\psi)^{c+1} = (x_\psi, x_\psi, x_\psi, x_\psi).
\end{equation}

\noindent This is a bottom-up procedure since granules start as a point and dilate thereafter. Subsequently, when $\textbf{x}(k+1)$ is available, a supervised learning step is given by considering the input-output pair, $(\textbf{x}(k),\textbf{x}(k+1))$, and the Recursive Least Squares method \cite{Leite2} \cite{Astrom1}. Thus,

\vspace{-12pt}

\begin{eqnarray} 
~~ \widetilde{A}^{c+1} = \hspace{-1pt} \left[ \begin{array}{cc}
1 \hspace{-2pt} & \textbf{0} \\
\textbf{a}_{0}^{c+1} \hspace{-2pt} & A^{c+1}      
\end{array} \right] ~ \textrm{and} ~~ 
\widetilde{B}^{c+1} = \hspace{-1pt} \left[ \begin{array}{c}
0 \\
B^{c+1}     
\end{array} \right],
\end{eqnarray}

\vspace{-4pt}

\noindent being the offset coefficients, $a_{\psi0}^{c+1} = x_{\psi}(k+1) / x_{\psi}(k)$, $\psi = 1, ..., \Psi$, initially. $B^{c+1} = [1 ~ 0 ~ ... ~ 0]^T$ is constant.

Updating a chosen $\bm{\mathcal{M}}^{i^*}$ consists in expanding the support $[l^{i^*}_\psi,L^{i^*}_\psi]$ and updating the core $[\lambda^{i^*}_\psi,\Lambda^{i^*}_\psi]$ of its components. Among all granules $\bm{\mathcal{M}}^i$ that can be expanded to include an $\textbf{x}(k)$, that with the highest similarity according to (\ref{sss}), $\bm{\mathcal{M}}^{i^*}$, in which

\vspace{-8pt}

\begin{equation}
    i^* = arg \max_{i = 1,...,c} (S(\textbf{x},\bm{\mathcal{M}}^i)),
\end{equation}

\vspace{-4pt}

\noindent is chosen. Granular coverage of the space of states and model memory to support online LMI fuzzy control design are emphasized by: (i) ignoring the deleting-by-inactivity and granule merging procedures of the original FBeM algorithm \cite{Leite1}; (ii) keeping model granularity, $\rho$, constant along the time steps; and (iii) considering the following SS-FBeM updating relations:

\vspace{-20pt}

\begin{equation} 
\begin{array}{llllllll} \textrm{If} \hspace{-6pt} & x_\psi(k) \in [L^{i^*}_\psi-\rho,l^{i^*}_\psi] & \hspace{-2pt} \textrm{then} ~~ l^{i^*}_\psi(k+1) = x_\psi(k)  \nonumber
\end{array}
\end{equation}

\vspace{-22pt}

\begin{equation}
\begin{array}{llllllll} \textrm{If} \hspace{-6pt} & x_\psi(k) \in [L^{i^*}_\psi,l^{i^*}_\psi+\rho] & \hspace{-2pt} \textrm{then} ~~ L^{i^*}_\psi(k+1) = x_\psi(k) \nonumber
\end{array}
\end{equation}

\vspace{-22pt}

\begin{equation} 
\begin{array}{llllllll} \textrm{Otherwise,} \hspace{-6pt} & l^{i^*}_\psi(k+1) = l^{i^*}_\psi(k) \textrm{ and} ~ L^{i^*}_\psi(k+1) = L^{i^*}_\psi(k) \nonumber
\end{array}
\end{equation}

\noindent for $\psi = 1, ..., \Psi$. Core parameters are updated from

\vspace{-17pt}

\begin{equation}
\lambda^{i^*}_\psi(k+1) = \Lambda^{i^*}_\psi(k+1) = \frac{l^{i^*}_\psi(k+1) + L^{i^*}_\psi(k+1)}{2}  \nonumber
\end{equation}

\vspace{-8pt}

The learning algorithm creates a new granule $\mathcal{M}^{c+1}$ or adapts the parameters of $\mathcal{M}^{i^*}$, accordingly. The Recursive Least Squares method updates $\widetilde{A}^{i^*}$ \cite{Leite2} \cite{Astrom1}.

\vspace{-2pt}

\section{Evolving LMI control design} \label{sec:lmi_control_design}

\vspace{-2pt}

From the state-space fuzzy model, we formulate conditions for Lyapunov stability and bounded control inputs as an LMI feasibility problem. The gain matrices of the fuzzy controller are derived automatically from the LMI whenever the SS-FBeM model changes.

\vspace{-1pt}

\subsection{Closed-loop control system}

\vspace{-1pt}

We consider parallel distributed compensation, i.e., the fuzzy controller and model rules share the same antecedent terms. The controller rule is

\vspace{3pt}

\noindent ~~ $R^i$: IF $x_1(k)$ is $\mathcal{M}_1^i$ AND ... AND $x_{\Psi}(k)$ is $\mathcal{M}_{\Psi}^i$

\noindent ~~~~~ \hspace{1.5pt} THEN $\widetilde{\textbf{u}}^i(k+1) = K^i \widetilde{\textbf{x}}(k)$

\vspace{3pt}

\noindent in which $\widetilde{\textbf{x}}(k) = [1 ~ x_1(k)$ ... $x_\Psi(k)]^T$ and  $\widetilde{\textbf{u}}(k) \hspace{-1pt} = \hspace{-1pt} [u_0(k)$ $u_1(k)$ \hspace{-4pt} ... $u_\Phi(k)]^T$. $K^i \in \Re^{(\Phi+1 \times \Psi)}$ is a gain matrix -- with offset term in the first column -- to be determined to make the closed loop system asymptotically stable, and/or to drive the states faster and smoother to a reference. Superscript $i$ on the left-hand side of the consequent state-feedback law means a local control input. The overall control signal is

\vspace{-15pt}

\begin{eqnarray}
~~ \widetilde{\textbf{u}}(k+1) = \sum\limits_{i=1}^c \mu^{ir} \widetilde{\textbf{u}}^i(k+1), \label{controlest}
\end{eqnarray}

\vspace{-3pt}

\noindent in which $\mu^{ir}$ is the re-scaled activation degree (\ref{boundc}).

Having model and controller $c$ rules, thus, Eqs. (\ref{outest}) and (\ref{controlest}) combined yield the closed-loop system,

\vspace{-18pt}

\begin{eqnarray}
~~ \widetilde{\textbf{x}}(k+1) = \sum\limits_{i=1}^c \sum\limits_{j=1}^c \mu^{ir} \mu^{jr} G^{ij} \widetilde{\textbf{x}}(k), \label{closedloop}
\end{eqnarray}

\vspace{-5pt}

\noindent in which $G^{ij} := \widetilde{A}^i + \widetilde{B}^i K^j$, or, equivalently,

\vspace{-10pt}

\begin{eqnarray}
\widetilde{\textbf{x}}(k+1) &=& \sum\limits_{i=1}^c (\mu^{ir})^2 G^{ii} \widetilde{\textbf{x}}(k) + \nonumber \\
&+& 2 \sum\limits_{i<j}^c \mu^{ir} \mu^{jr} \left(\frac{G^{ij} +  G^{ji}}{2}\right) \widetilde{\textbf{x}}(k).
\end{eqnarray}

\vspace{-5pt}

\noindent If the unforced system is stable, $K^i ~ \forall i$ may improve the transient response. Unstable systems require suitable $K^i$'s for stabilization primarily. An issue in evolving environment is that granules $\bm{\mathcal{M}}^i$, system matrices $A^i$, and the number of rules $c$, are time-varying. Local $K^i$'s should be reviewed after any model change.

\vspace{-1pt}

\subsection{Lyapunov stability conditions}

\vspace{-1pt}

A fuzzy Lyapunov function is a fuzzy combination of quadratic functions of the system states,

\vspace{-9pt}

\begin{equation}
V(\textbf{x}) = \sum_{i=1}^c \mu^{ir} \textbf{x}^T P^i \textbf{x}, \label{flf}
\end{equation}

\vspace{-4pt}

\noindent where $P^i > 0 ~ \forall i$. A stabilization result for the closed-loop system \eqref{closedloop} based on \eqref{flf} is as follows \cite{Feng1}.

\vspace{3pt}

\emph{Result}: The system \eqref{closedloop} is 
asymptotically stable if there exist positive definite matrices $X^i = (P^i)^{-1}$ and matrices $Q^i$ and $Z^i$, $i = 1,...,c$, such that

\vspace{-7pt}

\begin{eqnarray}
\left[ \begin{array}{cc}
X^i \hspace{-1pt} - \hspace{-1pt} (Z^j)^T \hspace{-2pt} - \hspace{-1pt} Z^j ~ \hspace{-2pt} & \hspace{-2pt} ~ (Z^j)^T \hspace{-1pt} (A^i)^T \hspace{-3pt} + \hspace{-2pt} (Q^j)^T \hspace{-1pt} (B^i)^T \\[\medskipamount]
A^i Z^j + B^i Q^j ~ & ~ -X^k
\end{array} \right] < 0 \label{LMI2}
\end{eqnarray}

\vspace{-1pt}

\noindent holds true for all combinations of $i,j,k = 1,...,c$. See \cite{Feng1} for a proof. The controller gains are

\vspace{-13pt}

\begin{eqnarray}
K^j = Q^j (Z^j)^{-1}, ~ j = 1,...,c. \label{gainmatrices}
\end{eqnarray}

\vspace{-4pt}

The number of fuzzy rules affects the complexity of LMI analyses. Finding a Lyapunov function for a large number of rules may be difficult \cite{Leite2} \cite {Tanaka1}. Due to aspects of the SS-FBeM learning algorithm, namely, inactive rules do not change, recalculation of gains \eqref{gainmatrices} is needed only for the active rules at a time step. Therefore, the number of LMIs in \eqref{LMI2} can be greatly reduced by considering active rules only.

\vspace{1pt}

\emph{Definition}: The number of active rules, $\mathfrak{x}$, for an instance $\textbf{x}(k)$ is equal to the number of terms that make the activation degree $\mu^{ir}(\textbf{x}(k)) > 0$, $i = 1, ..., c$.

Additionally, we reduced the number of concatenated rows in \eqref{LMI2} by making matrices $Z^j ~ \forall j$ and $X^k ~ \forall k$ equal to $X^i ~ \forall i$, with $X^i$ symmetric, to obtain a theorem.

\emph{Theorem}: The closed-loop system \eqref{closedloop} is 
asymptotically stable if there are positive definite matrices $X^i = (P^i)^{-1}$, and matrices $Q^j$; $i,j = 1, ..., \mathfrak{x}$, such that

\vspace{-15pt}

\begin{eqnarray}
\left[ \begin{array}{cc}
-X^i ~ & ~ X^i (A^i)^T \hspace{-2pt} + \hspace{-2pt} (Q^j)^T (B^i)^T \\[\medskipamount]
A^i X^i + B^i Q^j & -X^i
\end{array} \right] < 0 \label{LMI1}
\end{eqnarray}

\vspace{-6pt}

\noindent are satisfied for all combinations of $i,j = 1,...,\mathfrak{x}$ -- being $\mathfrak{x}$ the number of active rules for an instance $\textbf{x}(k)$. 

If a feasible solution is found, then the controller gains assigned to active rules are redefined as

\vspace{-16pt}

\begin{eqnarray}
K^i = Q^j P^i, ~~ i,j = 1,...,\mathfrak{x}. \label{gainsquare}
\end{eqnarray}

\vspace{-6pt}

\noindent The gains related to inactive rules are kept the same, as computed in previous time steps.

The feasibility problem \eqref{LMI1} is dynamic and convex. Finding a solution means that \eqref{flf} is Lyapunov for $i = 1,...,\mathfrak{x}$; and \eqref{closedloop} is stable using $K^i$ \eqref{gainsquare}. The proof of the Theorem follows analogously to that in the appendix of \cite{Leite2}. Efficient LMI parser and solver, Yalmip'18 \cite{Lofberg}, and Mosek'20 \cite{Andersen}, are available. 


\vspace{-3pt}

\subsection{Bounded control input}

\vspace{-3pt}

Being the current state, $\textbf{x}(k)$, known, the following result applies for a bounded control input \cite{Tanaka1}.

\emph{Result}: Given positive definite matrices $X^i$, and matrices $Q^j = K^i X^i$, $i = 1, ..., \mathfrak{x}$, as in \eqref{gainsquare}. The constraint $||\textbf{u}(k+1)||_2 \leq \zeta$ is enforced if

\vspace{-14pt}

\begin{eqnarray}
\left[ \begin{array}{cc}
1 & \textbf{x}(k)^T \\
\textbf{x}(k) & X^i 
\end{array} \right] > 0 ~
\textrm{ and } \left[ \begin{array}{cc}
X^i & (Q^j)^T \\
Q^j & \zeta^2 I 
\end{array} \right] > 0
\label{LMIxxxx}
\end{eqnarray}

\vspace{-6pt}

\noindent hold true for $i,j = 1,...,\mathfrak{x}$. See \cite{Tanaka1} for a proof. We replaced the initial state $\textbf{x}(0)$ in \eqref{LMIxxxx} by the current state $\textbf{x}(k)$ for online design. Parameter $\zeta$ bounds the maximum input, thus keeping it within the operation range of actuators. Eqs. \eqref{LMIxxxx} are appended to \eqref{LMI1} for stable fuzzy controllers satisfying input constraints.

\vspace{-3pt}

\section{Results on Evolving Modeling and Control of the Henon Nonlinear Map} \label{sec:results}

\vspace{-3pt}

The effectiveness of the SS-FBeM learning and model-based control is evaluated from a deterministic chaotic system, the Henon map \cite{Henon}. We use the Henon equations to generate a data stream. Evolving state-space fuzzy modeling and control are performed on the fly.

\vspace{-4pt}

\subsection{Henon map}

\vspace{-3pt}

The nonlinear equations of the Henon map are:

\begin{eqnarray}
x_1(k+1) \hspace{-3pt} &=& \hspace{-3pt} x_2(k) + 1 - \alpha x_1(k) x_1(k) \nonumber \\
x_2(k+1) \hspace{-3pt} &=& \hspace{-3pt} \beta x_1(k). \label{hen1}
\end{eqnarray}

\vspace{-8pt}

\begin{figure}[h]
\begin{center}
\includegraphics[width=.99\columnwidth]{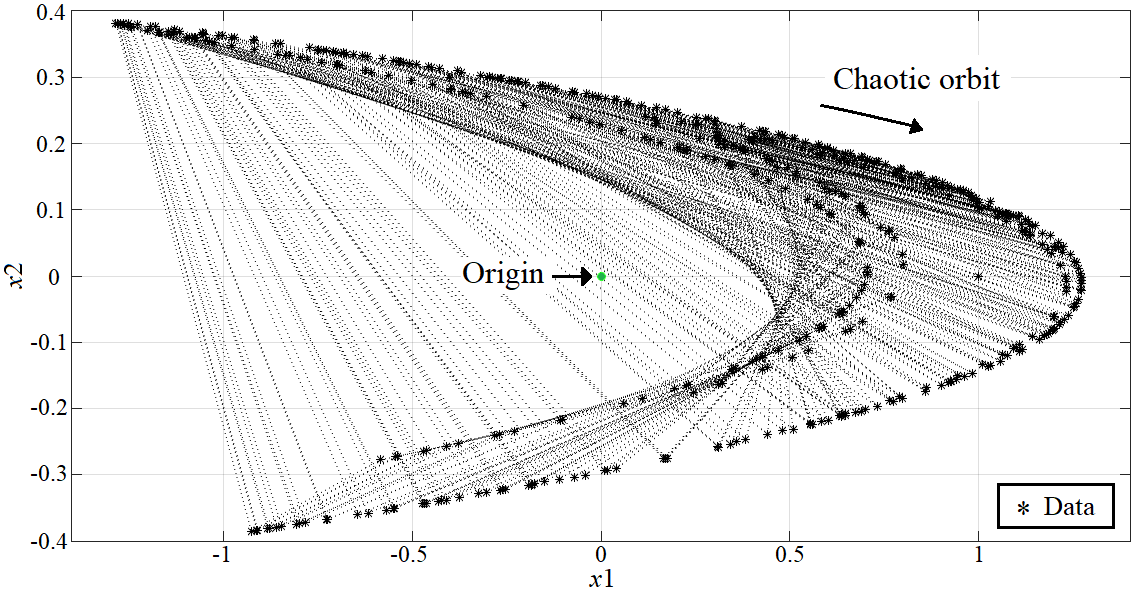}
  \vspace{-20pt}
  \caption{\label{HenonMap} Henon chaotic map: phase plane.}
\end{center}
\end{figure}

\noindent The phase portrait for $\alpha = 1.4$, $\beta = 0.3$, and initial state $\textbf{x}(0) = [1 ~~ 0]^T$ is shown in Figure \ref{HenonMap}. In this case, $[0.6314 ~~ 0.1894]$ and $[-1.1314 ~~ -0.3394]$ are the fixed points. If the orbit spreads over the phase plane, then we have a stochastic process. As we see a deterministic curve, then we have chaos. In fact, the Henon map is a model of the Poincaré section of the Lorenz system. Notice that the orbit of the unforced system settles into an irregular oscillation, confined in a fractal set, which never repeats exactly. 

\vspace{-5pt}

Aiming at stabilizing \eqref{hen1}, i.e., driving the chaotic orbit $\textbf{x}(k)$ to the origin $\textbf{x}^* = [0 ~~ 0]$, $\mathrm{lim}_{k\rightarrow\infty} (\textbf{x}(k) - \textbf{x}^*) = 0$, we add a control input, $u(k)$. Thus,

\vspace{-13pt}

\begin{eqnarray}
x_1(k+1) \hspace{-3pt} &=& \hspace{-3pt} x_2(k) + 1 - 1.4 x_1(k) x_1(k) + u(k) \nonumber \\
x_2(k+1) \hspace{-3pt} &=& \hspace{-3pt} 0.3 x_1(k). \label{hen2}
\end{eqnarray}

\vspace{-5pt}

The evolving model-based fuzzy controller is designed to opportunely inject values $u(k)$ to lead the states $\textbf{x}(k)$ to the origin $\textbf{x}^*$. The Henon system \eqref{hen2} is completely controllable and observable.

\vspace{-2pt}

\subsection{One-step SS-FBeM prediction}

\vspace{-2pt}

Consider the evolving fuzzy controller off for $k_f$ time steps, i.e., $u(k) = 0, ~ k = 1,...,k_f$. We aim to evaluate the SS-FBeM accuracy in one-step prediction. The model is built from scratch, with no knowledge about the dynamic system that generates the data stream. 

Let the root mean square error be

\vspace{-12pt}

\begin{eqnarray}
\textrm{RMSE} = \frac{1}{k_f} \sum_{k=1}^{k_f} \sqrt{ \sum_{j=1}^{n} \left(x_j(k+1)-\tilde{x}_j(k+1)\right)^2}, \label{rmsee}
\end{eqnarray}

\vspace{-4pt}

\noindent in which $\textbf{x}(k+1)$ is the actual value given by \eqref{hen1}; and $\tilde{\textbf{x}}(k+1)$ is the SS-FBeM estimate \eqref{outest}. Table \ref{tab1} shows the one-step prediction results for $k_f = 500$, and different maximum width $\rho$ allowed for granules.

\begin{table}[h]
\begin{center}
\caption{SS-FBeM modeling and one-step prediction results due to different granularities $\rho$.}
\vspace{1ex}
\begin{tabular}{c|cc} \hline
    Granularity ($\rho$) & Structure ($\#$ rules) & RMSE \\\hline
    0.1 & 60 & 0.0305 \\
    0.2 & 32 & 0.0317 \\
    0.3 & 20 & 0.0359 \\
    0.4 & 16 & 0.0442 \\
    0.5 & 10 & 0.0445 \\
    0.6 & 9 & 0.0478 \\
    0.7 & 7 & 0.0487 \\
    0.8 & 5 & 0.0573 \\
    \hline
\end{tabular}
\label{tab1}
\end{center}
\end{table}

Strictly speaking, we notice from Table \ref{tab1} that the number of fuzzy rules and, therefore, the number of SS-FBeM parameters, reduces as we increase the hyperparameter $\rho$, which facilitates further LMI control design. However, the SS-FBeM accuracy reduces so that the control is designed for a worse picture of the actual system, which may lead worse-than-expected transient responses or instability. A trade-off between model accuracy and compactness is noted. 

\begin{figure*}[!t]
\begin{center}
\includegraphics[width=1.99\columnwidth]{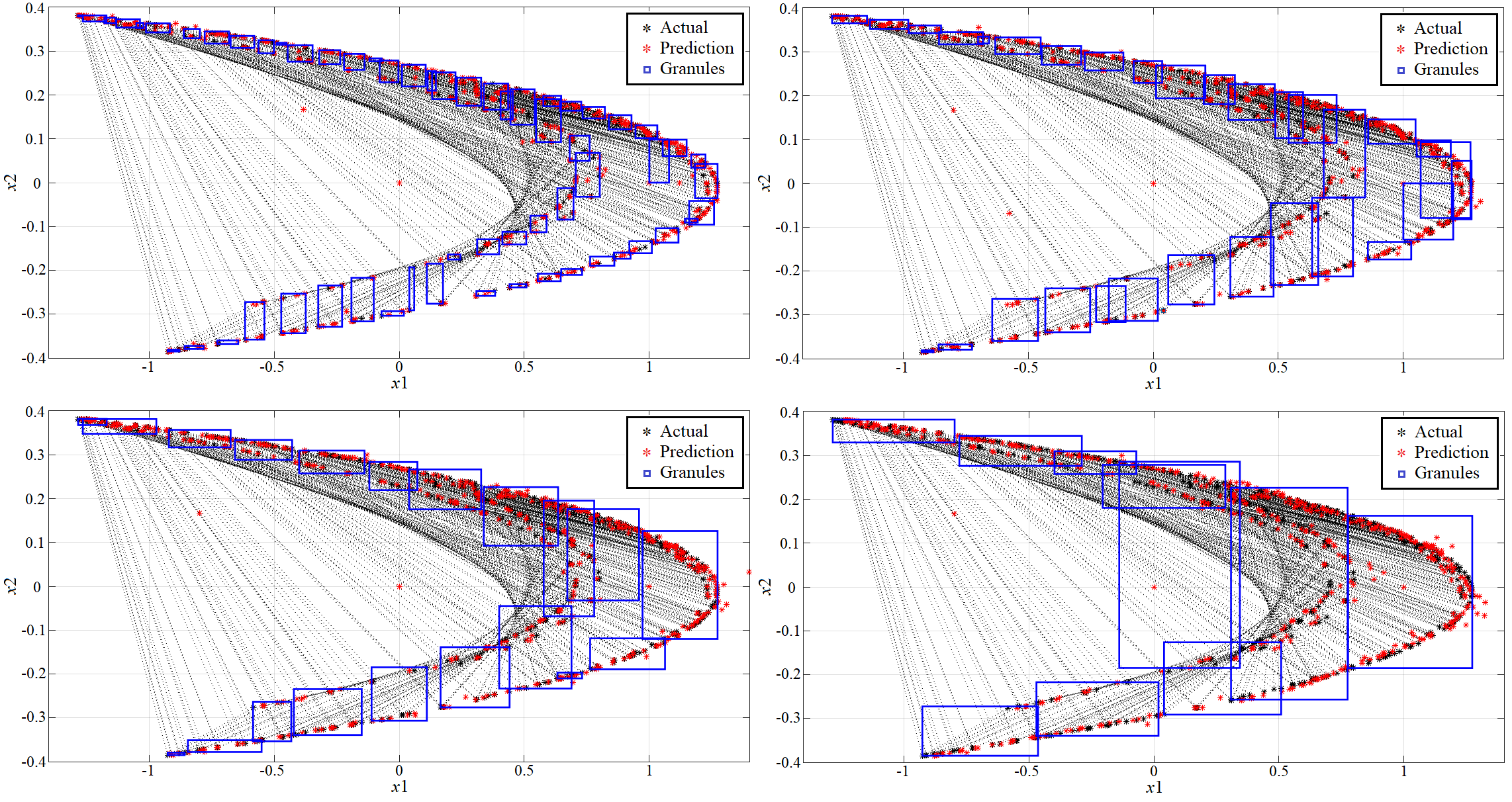}
  \vspace{-6pt}
  \caption{\label{figure3}SS-FBeM online incremental partition of the Henon state space using $60$, $32$, $20$, and $10$ granules.}
\end{center}
\end{figure*}

\begin{figure}[ht]
\begin{center}
\includegraphics[width=.99\columnwidth]{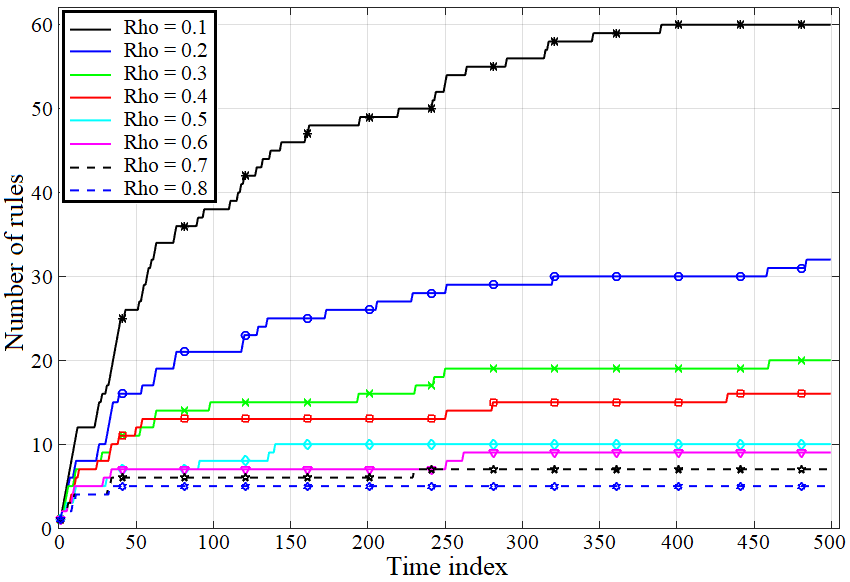}
  \vspace{-14pt}
  \caption{\label{figure4}SS-FBeM online structural evolution for different granularities $\rho \in [0.1, 0.8]$.}
\end{center}
\end{figure}

Figure \ref{figure3} shows how the state space is granulated for $\rho = 0.1, 0.2, 0.3, 0.5$, yielding more refined or coarser partitions. Figure \ref{figure4} shows the SS-FBeM online structural evolution for all cases. Naturally, the smaller the granules, the more precise the local model, but the more complex the LMI to be solved.

\subsection{Closed-loop stabilization}

The fuzzy controller is designed based on the evolving fuzzy model. Nonetheless, the control input is applied to the original Henon system \eqref{hen2}. Thus $\textbf{x}$ is the actual system state, i.e., we have controlled \textit{unknown} chaotic systems on the fly.

Define the settling time, $t_s$, as the number of time steps from the application of the control input, $u$, until all states, $\textbf{x}$, enter and remain within a 2\% band around the origin, $\textbf{x}^*$. The range of values of the system \eqref{hen1} are $x_1 \in [-1.2836, ~ 1.2727]$ and $x_2 \in [-0.3851, ~ 0.3818]$. Thus, $x_1(k) \leq x_{1(2\%)} =  0.0256$, and $x_2(k) \leq x_{2(2\%)} = 0.0077$, $\forall k$, after $u \neq 0$, determines $t_s$.

Let the energy $E_{\textbf{x}}$ of the states $\textbf{x}$, for $u \neq 0$, be

\vspace{-12pt}

\begin{equation}
E_{\textbf{x}} = \langle \textbf{x}(k), \textbf{x}(k) \rangle = \sum_{j=1}^{n}{\sum_{k_{u \neq 0}}{|x_j(k)|^{2}}}.
\end{equation}
\noindent In particular, $n=2$ is the number of states. We turned the fuzzy controller on and off at $k=500$ and $k=600$. The lower the values of $t_s$ and $E_{\textbf{x}}$, the better. Table \ref{tab2} summarizes the results for different granules sizes, $\rho$, and maximum control input, $\zeta$. 

\begin{figure}[ht]
\begin{center}
\includegraphics[width=.99\columnwidth]{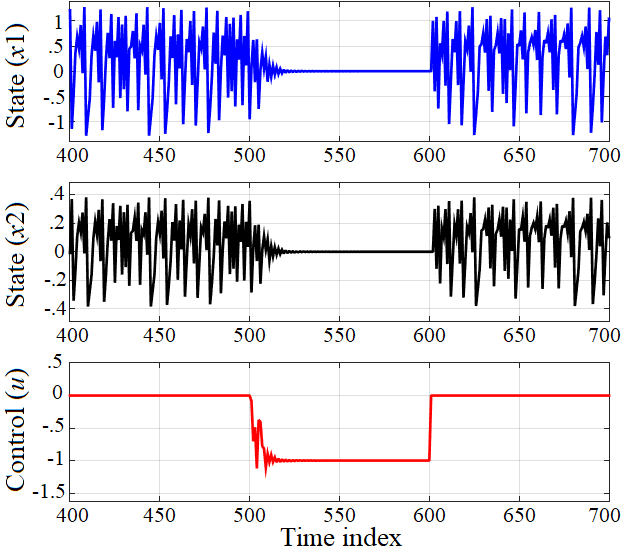}
  \vspace{-15pt}
  \caption{\label{figureControl}Stabilizing control: SS-FBeM, with $\rho = 0.2$, learns constantly; the fuzzy controller, with $\zeta = 1.5$, is designed and updated in real-time, from $k = 500$ to $k = 600$, to regularize the Henon chaos.}
\end{center}
\end{figure}

\begin{figure}[ht]
\begin{center}
\includegraphics[width=.99\columnwidth]{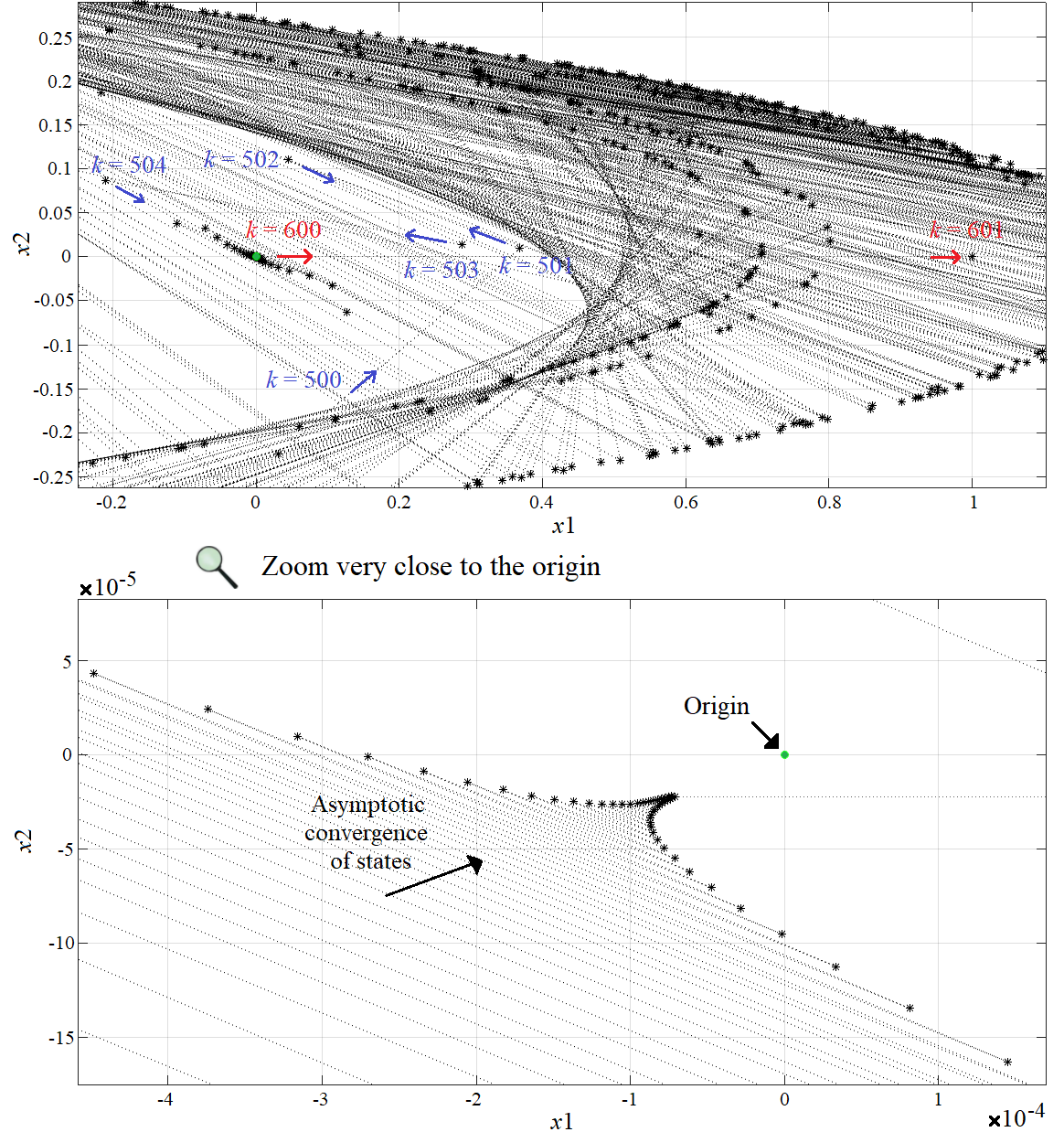}
  \vspace{-20pt}
  \caption{\label{figure6}State-space perspective of the result in Fig. \ref{figureControl} elucidating the convergence of the Henon states, which spiral towards the origin. Control is applied at $k = 500$ (blue arrows). The system backs to the chaotic orbit at $k = 600$ when the control is turned off (red arrows).}
\end{center}
\end{figure}

\begin{table}[h]
\begin{center}
\caption{\label{tab2}SS-FBeM control results.}
\vspace{1ex}
\begin{tabular}{cc|cc} \hline
    Gran $\rho$ & Bound $\zeta$ & Settling [$k$] & Energy $E_{\textbf{x}}$ \\\hline
    0.1 & 1.5 & 21 & 1.9574 \\
    0.2 & 1.5 & 23 & 1.9629 \\
    0.3 & 1.5 & \textbf{17} & \textbf{1.5953} \\
    0.4 & 1.5 & 85 & 1.9152 \\
    0.5 & 1.5 & 39 & 1.9131 \\
    0.6 & 1.5 & 61 & 2.7607 \\
    0.1 & 1.1 & 53 & 2.1995 \\
    0.2 & 1.1 & 53 & 2.1264 \\
    0.3 & 1.1 & 77 & 2.0304 \\
    0.4 & 1.1 & 46 & 2.4073 \\
    0.5 & 1.1 & -- & 2.5735 \\
    0.6 & 1.1 & -- & 3.8334 \\
    \hline
\end{tabular}
\end{center}
\end{table}
We notice from Table \ref{tab2} that an intermediate granularity, $\rho = 0.3$, and a wider amplitude range for inputs, $\zeta = 1.5$, provide the best closed-loop performance. As expected, larger values of $\rho$ cause higher modeling error, such that the control is designed based on a less accurate model. Therefore, in cases such as $\rho = 0.5, 0.6$, after $100$ time steps, despite converging evidence, the states still spiral on a band larger than $2\%$ around $\textbf{x}^*$. To exemplify the regularization of the chaotic behavior, Figure \ref{figureControl} shows the states convergence when $u$ is enabled at $k = 500$. Figure \ref{figure6} highlights the asymptotic convergence from the phase plane perspective. Notice that the Henon system backtracks to its chaotic orbit when the control is turned off at $k = 600$.

\vspace{-8pt}

\section{Conclusion} \label{sec:conclusion}

\vspace{-8pt}

This study describes a state-space variation of fuzzy evolving method to model and control unknown nonlinear dynamical systems. SS-FBeM evolves the structure and parameters of a fuzzy model and a fuzzy controller from scratch. The incremental learning method is particularly devoted to pave the data space and keep memory of past experiences. Control gains are derived from an LMI feasibility problem to guarantee closed-loop Lyapunov stability and bounded inputs. 

We have shown asymptotic stabilization of the Henon chaos as application example. The results remark a low RMSE in one-step prediction, 0.0359; and a settling time (within 2\%) of 17 steps using the original Henon equations in the loop and an intermediate SS-FBeM granularity, 0.3. The Henon equations are assumed unknown for any modeling and control design purpose. SS-FBeM perceives the dynamic system by means of the data stream only. We envision applications on secure communication, suppression of interference and artifacts in bio-signals, and control of time-varying systems in general. In the future we will address trajectory following control by suppressing chaos.


\bibliographystyle{eusflat2021}
\bibliography{BIBeusflat2021}

\end{document}